\documentstyle[12pt]{article}
\def\beq{\begin{eqnarray}}
\def\eeq{\end{eqnarray}}
\def\bea{\begin{eqnarray*}}
\def\eea{\end{eqnarray*}}
\def\be{\begin{equation}}
\def\ee{\end{equation}}

\def\centeron#1#2{{\setbox0=\hbox{#1}\setbox1=\hbox{#2}\ifdim
\wd1>\wd0\kern.5\wd1\kern-.5\wd0\fi
\copy0\kern-.5\wd0\kern-.5\wd1\copy1\ifdim\wd0>\wd1
\kern.5\wd0\kern-.5\wd1\fi}}
\def\ltap{\;\centeron{\raise.35ex\hbox{$<$}}{\lower.65ex\hbox{$\sim$}}\;}
\def\gtap{\;\centeron{\raise.35ex\hbox{$>$}}{\lower.65ex\hbox{$\sim$}}\;}

\newcommand{\newc}{\newcommand}
\newc{\qbar}{{\overline q}}
\newc{\Kahler}{K\"ahler }
\newc{\deltaGS}{\delta_{\rm GS}}

\begin{document}

\begin{titlepage}

\vskip 1.2cm \vspace*{1cm}

\begin{center}

{\LARGE{\bf The Supersymmetric Axion and Cosmology}}

\vskip 1.4cm

{\large A. Yu. Anisimov}
\\
\vskip 0.4cm {\it Institute of Theoretical and Experimental
Physics, Moscow }
\\

\vskip 4pt

\vskip 1.5cm

\begin{abstract}

In this lecture\footnote{This lecture was given at XII'th Baksan
International School of Particles and Cosmology, 21-27 April 2003,
Baksan Valley, Kabardino-Balkaria.} we review several cosmological
issues associated with the axion. Axion solves the strong CP
problem and is a good candidate for the dark matter. Limits, which
are imposed by the value of isocurvature fluctuations fraction in
the observed CMBR from WMAP data and the domain wall problem are
discussed in a supersymmetric flat direction axion model.

\end{abstract}

\end{center}

\vskip 1.0 cm
\end{titlepage}
\section{Introduction}

At present astrophysical data suggest that the observable part of the
Universe contributes only a few percents to the total energy density.
 Most of the content of
the Universe is in the form of the dark matter (25$\%$) and dark
energy (70$\%$). In this lecture we will review one of the most
viable candidate for the dark matter, which is  axion
particle,\footnote{There are two other candidates for the dark
matter, which are neutralino or LSP and mirror world matter, which
we do not discuss here.} that originally arose as a solution to
the strong CP problem. Axion decay constant must be very large
compared to the electroweak scale, in order to not confront
astrophysical data, which yields  the so-called ''invisible axion"
models. Since axion interacts with other particles with the
strength  inversely proportional to its decay constant it turns to
be a good dark matter candidate.

It is worth to mention that string theory provides many periodic
fields, associated with global symmetries spontaneously broken
down to some discrete subgroups. Although, naively the decay
constants of string axions are of order of ${M_{pl}\over
32\pi^2}\sim 10^{16}{\rm GeV}$, which is too large for a QCD axion
decay constant, one might hope to construct a QCD axion model out
of string theory axions. The rest of these periodic fields may
yield a solution to the dark energy if they get stuck away from
the minimum of their potentials.

Below we discuss some axion models stressing some important issues
which arise if the QCD axion does exist and contribute to the dark
matter. In section 2 we briefly remind the usual QCD story and the
originally proposed ''invisible" axion models. In the section 3 we
discuss supersymmetric axion which corresponds to some flat
direction. Finally we discuss axion isocurvature fluctuations and
the domain wall problem in the context of the supersymmetric axion
model. The last two issues impose a significant constraints on any
axion model. We review how one can avoid corresponding
difficulties.

\section{The Strong CP Problem and The ''Invisible" Axion.}
The Lagrangian of QCD, besides ordinary terms, contains the total
derivative term or $\theta$-term:
\be
{\mathcal L}_{\theta}=\frac{\theta}{32\pi^2}G_{\mu\nu}^a\tilde
G^{a\mu\nu},
 \ee
 where $\tilde G^{a\mu\nu}$ is the dual to the field strength
 tensor. This term is gauge and Lorentz invariant, therefore it
 should be incorporated in the  full QCD Lagrangian. However, one
 can check easily that this $\theta$-term is a  total
 derivative since it can be rewritten as
 \be
{\mathcal L}_{\theta}=\theta\partial_{\mu}K^{\mu}
 \ee
 with
 \be
 K^{\mu}= \frac{1}{16\pi^2}\epsilon^{\mu\nu\alpha\beta}(A^a_{\nu}\partial_{\alpha}A_{\beta}^a+
 \frac{1}{3}f_{abc}A_{\nu}^a A_{\alpha}^b A_{\beta}^c).
 \ee
This term being a total derivative does not affect the equation of
motion and if one would have trivial vacuum structure like that in
QED this term would not affect any physics at all. However,
because of the rather complicated vacuum structure of QCD, i.e.
the presence of instantons, the integral over this total
derivative does not vanish. Therefore, this term contributes to
the action of QCD and must be retained.

First, let us note that this term violates CP symmetry, thus it
can give rise to neutron electric dipole moment. Current
experiments set a strong bound on the value of
$\theta$\footnote{Neutron electric dipole moment can be computed
using current algebra methods \cite{edm}.}, which is
\be
\theta < 10^{-9}. \ee

It has to be explained why this number is so unnaturally small.
Peccei and Quinn \cite{pq} proposed a very elegant solution to
that problem (although not unique). Namely, consider the case when
the Lagrangian has an additional $U(1)_{PQ}$ symmetry. If this
symmetry is broken at some energy scale, which we denote by $f_a$,
there will be NG-boson generated. This NG-boson is called
axion\footnote{One of the other two known solutions to the strong
CP-problem is what is called Nelson-Barr
mechanism\cite{nelsonbarr}. Second possibility  is to set the mass
of $u$-quark to be zero.}. It is massless  and it couples to the
gauge bosons and to the matter fields with the strength which is
inversely proportional to the scale of the PQ-symmetry breaking
$f_a$. Nonperturbative instanton QCD effects, however, give an
axion a potential of the form, which is well approximated by
\be
V_{QCD}(T)=f_a^2m_a^2(T)[1-\cos\left(\frac{aN}{f_a}\right)],
\label{pot} \ee breaking $U(1)$ down to discrete subgroup $Z_N$,
with the mass of the axion depending on the quark masses and on
the temperature of the quark--gluon plasma. This naturally sets
the coefficient in front of axion-gluon-gluon term in the QCD
Lagrangian to zero, since minimum of that potential is at $a=0$.
Thus, PQ solution solves the strong CP problem.

 At zero
temperature the axion mass is given by
\be
m_a^2=\frac{m_u m_d m_s}{m_u m_d + m_d m_s + m_s
m_u}\frac{\Lambda^3_{QCD}}{f_a^2}\sim
\frac{f_{\pi}^2m_{\pi}^2}{f_a^2}, \ee

where $\Lambda_{QCD}$ is the QCD scale and can be related to the
pion mass and decay constant by
$m_{\pi}^2=2(m_u+m_d)\Lambda^3_{QCD}/f_{\pi}^2$. If $\Lambda_{QCD} \ll
f_a$ then the axion turns out to be very light comparing to the
QCD scale. At high temperature axion is even lighter and its mass
is temperature dependent \cite{turner}
\be
m_a(T)=0.1m_a(T=0)\left(\frac{\Lambda_{QCD}}{T}\right)^{3.7}. \ee

There are   three historically consequential models of the axion :
\begin{itemize}
\item Weinberg-Wilczek axion

\item KSVZ axion (Kim; Shifman, Vainshtein, Zakharov) \cite{ksvz}

\item ZDFS axion (Zhitnitskii; Dine, Fishler, Srednicki) \cite{zdfs}
\end{itemize}
In the first model $U(1)_{PQ}$ symmetry is broken at the  EW
scale. There are two Higgs doublets and the axion is
\be
a=\frac{1}{v}(v_{\phi}Im \phi_0 - v_{\chi}Im \chi_0), \ee where
$\phi_0$ and $\chi_0$ are the neutral components of the Higgs
doublets. The PQ-symmetry is broken at
$v=\sqrt{v_{\phi}^2+v_{\chi}^2}$, where $v\approx 250 {\rm GeV}$.
Such an axion is already ruled out by the experiment because it
would lead to disastrous loss of energy by various cosmological
objects via axion emission. The other constraint which is even
stronger comes from accelerator experiments.

The second model and the third are so-called ''invisible" axion
models. In the KSVZ model one introduces a complex scalar field
$\Phi$ which couples to the hypothetical quark field in the
fundamental representation of $SU(3)_c$:
 \be
\delta {\mathcal L}=\Phi \bar Q_R Q_L +h.c.
 \ee
$\Phi$ is supposed to develop large expectation value
$f_a/\sqrt{2}$ ($f_a\gg \Lambda_{QCD}$) and axion is defined to be
 \be
a=f_a{\rm Arg}\Phi.
 \ee
The low-energy coupling of the axion to gluons is then given by
\be
\delta {\mathcal L}=
\frac{1}{32\pi^2}\frac{a}{f_a}G^a_{\mu\nu}\tilde G^{a\mu\nu},
 \ee
such that $\theta\rightarrow\theta
+\frac{a}{f_a}$ in the lagrangian and one may apply Peccei
and Quinn solution.

In the ZDFS model one starts with the additional scalar SM singlet
field $\Sigma$. The scale of PQ-symmetry breaking is separated
from the EW scale and the axion field is like in the
Weinberg-Wilczek model but with $\Sigma$ field added:
\be
a=\frac{1}{V}(v_{\phi}Im\phi_0-v_{\chi}Im\chi_0+v_{\Sigma}Im\Sigma_0),
 \ee
 where $V=\sqrt{v_{\phi}^2+v_{\chi}^2+v_{\Sigma}^2}\approx
 v_{\Sigma}$ if $v_{\Sigma}$ is large compared to the Higgs
 expectation value. It can be as large as GUT or even Planck
 scale.

Last  two models are minimal extensions of the standard model, in
which one can generate effectively the required coupling of the
axion to gluons. At the same time in both cases one has to
introduce a new scale, which is not defined by the model. This
scale is essentially time independent and constrained by
astrophysical and cosmological data to be in the range $10^{9}{\rm
GeV }<f_a<10^{12}{\rm GeV}$. In the next section we discuss a
model where one can go a little further then just introducing a
new scale, but rather relating $f_a$ to other parameters, such as
SUSY breaking scale and the Hubble constant. This would have an
important consequences, which are discussed in the next two
sections.

\section{Supersymmetric Case}

All previous models require some extension of the electroweak
model. However, one may want to consider some supersymmetric
version of the original picture. Such version would be, for
example, an NMSSM model, which  also yields a solution to $\mu$
problem in the MSSM. One adds additionally at least one singlet
chiral field with the most general superpotential. For a
particular choice of couplings this looks like supersymmetrised
version of previously mentioned ZDFS axion model. One can generate
axion-gluon-gluon term and apply PQ solution. However,
supersymmetric case is different. Besides light axion one has
additionally another PQ charged scalar, which is the radial part
of a complex PQ field. This partner of axion is called {\it
saxion}. Saxion acquires large mass due to supersymmetry breaking
and may influence evolution of the axion field. There are flat
directions associated with different fields and one can imagine
the following potential\footnote{This is similar to \cite{lyth1}.}
\be
V(S)=\frac{\lambda |S|^6}{M_{pl}^2}-H^2|S|^2-
m_{3/2}^2|S|^2+V_{QCD}(T), \ee where $H$ is the Hubble constant,
$S=|S|{\rm exp}(a/|S|)$ is the PQ superfield, $|S|$ is saxion, and
$a$ is axion. Such potential naturally appears as a result of
lifting of the flat direction $S$ by
\begin{itemize}

\item supersymmetry breaking effects in the early Universe, which
leads to the masses of order of $H^2$ \cite{diner},

\item nonrenormalizable operators, suppressed by the Planck scale,
the lowest of which we take as an example in the
potential above,

\item zero curvature term of order of the SUSY breaking scale.
\end{itemize}

This potential has a form of of the ''mexican hat", and
corresponding $U(1)$ symmetry is broken at the energy scale
\be
f_a=|S|\sim (HM_{pl})^{\frac{1}{2}} \ee at early time. Note, that
this effective scale of $U(1)$ breaking is time--dependent until
$H\sim m_{3/2}$, when it gets frozen at $f_a\sim
(m_{3/2}M_{pl})^{1/2}\sim 10^{11}{\rm GeV}$. We will discuss
simple model with one inflaton which decays in the end of
inflation quickly reheating the Universe up to some temperature
$T_R$. In that case Universe becomes first radiation dominated.
While it expands, it cools and at temperature $T\leq
\Lambda_{QCD}$ the NG boson, which travels along the bottom of
''mexican hat'' gets the mass $m_a(T\leq \Lambda_{QCD})\approx
f_{\pi}m_{\pi}/f_a$ and starts to oscillate. This happens at the
value of the Hubble constant given by
\be
H_{QCD}=\left(\frac{\Lambda_{QCD}}{T_R}\right)^2H_I, \ee where
$H_I$ is the value of the Hubble constant during inflation. Taking
$T_R\sim 10^{9}{\rm GeV}$, and $H_I\sim 10^{13}{\rm GeV}$ one
obtains
\be
H_{QCD}\sim 10^{-5}{\rm GeV}. \ee The initial amplitude of the
axion field is $a_0\sim f_a$, which corresponds to
$\frac{a}{f_a}=\theta_a\sim {\mathcal O}(1).$\footnote{There is no
reason why initially the phase of the axion should be  much
different from unity, unless one employs anthropic principle.}
Therefore from the kinetic term one can get a rough estimate of
the axion energy density at this moment
\be
\rho\sim m_a^2f_a^2\approx f_{\pi}^2m_{\pi}^2\approx (0.1 {\rm
GeV})^4, \ee which constitutes a tiny fraction of the total energy
density
\be
\frac{\rho_a}{\rho_{tot}}\sim \frac{f_a^2}{M_{pl}^2}\approx
10^{-16}.\ee
 More accurate estimates, including decays of the
axionic strings and domain walls lead to a larger value by an
order of magnitude (for review, see, for example
\cite{lyth2,japan}). Axion is so light and weakly interacting that
it remains being unthermalized, coherently oscillating field until
present times contributing to the CDM density.

Since axion exists in the form of the cold matter its energy
density decreases slower by a factor of $a$, then that of
radiation, with $a$ being the scale factor in Robertson-Walker
metric. Because of that, this small fraction grows over time and
eventually starts to dominate the energy density. In the simple
model, which we consider, that restricts $f_a$ to be somewhere
lower than $10^{11}-10^{12}{\rm GeV}$, which is consistent with
the value we have obtained above.

\section{Isocurvature Fluctuations and Domain Walls}

 First issue which shall be discussed is the limit coming from the
 observed CMB anisotropy \cite{wmap} on the value of isocurvature
 fluctuations produced by axion. Isocurvature
 fluctuations appear when there are two or more fields, which generate quantum fluctuations
 during inflation.
 Fluctuations in the density of one of them may be compensated  by some other field,
 so that curvature remains unperturbed. These kind of fluctuations
 thus are called {\it isocurvature} fluctuations.
 In the case of massless
 (or nearly massless) axion and massive saxion isocurvature fluctuations are
 fluctuations in PQ charge or, in other words, fluctuations in
the
 phase $\theta$. The size of primordial spectrum of such
 fluctuations is set by
 \be
 \frac{\delta\theta}{\theta}=\Omega_a\frac{H_I}{2\pi f_a},
 \ee
 where $\Omega_a$ is axion contribution to the total density.
This can be understood as follows. First, one finds that the
equation of motion
\be
\ddot\theta_k+3H\dot\theta_k+\frac{k^2}{a^2}\theta=0 \ee after
rescaling $\theta_k\rightarrow \frac{\tilde\theta_k}{a}$ and going
to conformal time reads as
\be
\tilde\theta_k''+(k^2-\frac{a''}{a})\tilde\theta_k=0, \ee and has
a solution
\be
\tilde\theta_k=\frac{e^{-ik\tau}}{\sqrt{2k}}(1+\frac{i}{k\tau}),
\ee which gives at superhorizon scales
\be
|\theta_k|^2=\frac{1}{2a^2 k}=\frac{H^2}{2k^3}.
 \ee
Since power spectrum is defined as
\be
{\mathcal P}(k)=\frac{k^3}{2f_a^2\pi^2}|\theta_k|^2, \ee one can
see that the size of the phase fluctuations is $H_I/2\pi f_a$.

The observed size of CMB anisotropy is of order of $10^{-5}$,
which means that
\be
\Omega_a\frac{H_I}{2\pi f_a}<10^{-5}\label{limit} \ee at
least\footnote{For more detailed analysis see \cite{lyth3}}.
Moreover, the observation tell us that fluctuations are almost
pure adiabatic which restricts (\ref{limit}) even stronger.
Although, it is well known that isocurvature fluctuations tends to
decay into adiabatic ones \cite{wands}, in the case of the axion
the limit above still is a considerable underestimate.

However, even (\ref{limit}) impose stringent constraints. Having
Hubble constant of order of $10^{12}{\rm GeV}$ requires $f_a>
10^{16}{\rm GeV}$, which is far above the upper limit on $f_a$
imposed  in order to not overclose the Universe. In the scenario
described in the previous section this problem is naturally
avoided, since $f_a$ is time dependent. During inflation its value
is given by $(H_IM_{pl})^{1/2}\sim 10^{16}{\rm GeV}$, which has
just about right size.\footnote{If one includes higher order
nonrenormalizable terms of the form $\frac{\lambda
|S|^{2n+4}}{M_{pl}^{2n}}$ the PQ symmetry is broken at higher
values $f_a>10^{16}{\rm GeV}$, which relaxes the constraint from
isocurvature fluctuations.} After inflation the value of $f_a$
decreases with time as $t^{-1/2}$ until it freezes at
$(m_{3/2}M_{pl})^{1/2}\sim 10^{11}{\rm GeV}$. Such scenario seems
quite plausible because it solves isocurvature fluctuations
problem, and one does not overclose the Universe at the same time.

The other question which shall be discussed is the domain walls
problem. The problem arises as following. When $U(1)$ is broken
down to $Z_n$, the vacuum of the theory falls into $N$ degenerate,
but disconnected and, thus, distinct regions. As it is well known,
this leads to topological defects which appears as kink solutions
between each of the disconnected piece of vacuum manifold. Simple
example is the Higgs--like potential of the real scalar field
which possess $Z_2$ discrete symmetry. Namely, this theory has two
vacua $\phi=\pm v$. If there are two regions in space lying in
different vacua, one can easily obtain kink solution of the
classical equation of motion, with boundary conditions
\be
\phi(x=-\infty)=-v,~~ \phi(x=\infty)=v. \ee It easy to see that
there is energy density between boundaries, which has maximum at
$x=0$, and its characteristic thickness would be inversely
proportional to the mass of the scalar. This object is a simple
example of a domain wall.

Topologically formed domain walls in case of the exact discrete
symmetry are absolutely stable. Since domain walls are
two-dimensional objects their energy density behaves as $1/a$,
while the Universe is expanding. This is much slower than that for
matter ($1/a^3$) and radiation ($1/a^4$), thus, leading to
cosmological disaster very fast. Since we do not observe our
Universe to be domain walls dominated, means that if they were
ever formed their density either was diluted up to cosmologically
safe densities, i.e. less then one per horizon, or there are no
exact discrete symmetries, so that domain walls collapse before
they dominate Universe.\footnote{One other issue is that domain
walls, living long enough, lead to a nongaussianity of CMB, which
is in direct contradiction with current observation \cite{wmap2}.
}

Below we will discuss both possibilities to avoid domain walls
problem. A simple resolution comes in the case when PQ phase
transition occurs before or during inflation. Parts of the
Universe sitting in different vacua grow into very large regions,
such that our Universe turns out to be living in one of those. We
need to make sure, that reheating of the Universe does not restore
$U(1)$ at any time after inflation, so that the Universe stays in
the same vacuum until QCD effects turn on. Provided both
conditions are satisfied, domain walls which might be formed at
$T\sim \Lambda_{QCD}$ are not stable topologically. They do not
separate different vacua and, thus, will decay.

 For that to work we have to require that
\be
f_a>H_I,T_R, \ee which seems plausible in the picture described in
the previous section, since at early times the value of $f_a$ is
\be
f_a\sim(H_IM_{pl})^{\frac{1}{2}}\sim 10^{16}{\rm GeV}, \ee when
one takes $H_I\sim 10^{12}{\rm GeV}$ and $T_R\sim 10^{9}{\rm GeV}$
as in the simplest chaotic inflation scenario. Therefore, the simple construction
described in the Section 2 works well to solve the domain wall problems too.

There is one more case one may want to look at. Similar to
(\ref{pot}) we can imagine the potential of the form
\be
V(S)=\frac{\lambda |S|^6}{M_{pl}^2}+H^2|S|^2-
m_{3/2}^2|S|^2+V_{QCD}(T). \ee In that case phase transition
occurs late, with
\be
f_a\sim (m_{3/2}M_{pl})^{\frac{1}{2}}\sim 10^{11}{\rm GeV}.\ee
This automatically solves isocurvature fluctuations problem, since
there are none generated during inflation. But one is still
confronted with the domain walls problem in that case. There is
still a possibility to made these walls to collapse. For that one
needs to break $Z_n$ symmetry, adding, for example, to the
potential
\be
\delta V=\mu^4\cos(\theta), \ee where $\mu$ is some scale.
However, if one uses the restriction from neutron electric dipole
moment from one hand, and require domain walls to collapse fast
enough to not became cosmologically dangerous from another, one
arrives to the following criterium
\be
\left(\frac{m_a}{M_{pl}}\right)<\left(\frac{\mu}{f_a}\right)^2<10^{-9/2}\left(\frac{m_a}{f_a}\right),
\ee which numerically gives
\be
10^{-5}{\rm GeV}<\mu <10^{-4}{\rm GeV}\ee up to factors of order
one. This seems  rather narrow window, which is hard to satisfy.
Although, there might be constructed a model, where these limits
are much broader \cite{holdom}.

In the end, we should note that the domain walls problem is safely
avoided, if the discrete group is simply $Z_1$. This case can be
constructed, and domain walls will not be topologically stable. Therefore,
 the issue with domain walls is significantly relaxed.

Finally, we should mention that there are other cosmological
scenarios, i.e. those where the Universe is matter dominated soon
after inflation \cite{dine}, due to the moduli oscillations, such
that Universe reheats several times. First is due to inflaton
decay, second and further ones are due to moduli decay.
 There might also be an issue of parametric resonance effects
\cite{yanagida}, if one imagine that  axion potential is very
flat. Such scenario solves isocurvature fluctuations problem.
However, it poses a domain wall problem, because after inflation
axion acquires large fluctuations due to the instability, which is
similar to that in Mathieu equations. This case is safe from the
domain walls problem only in $N=1$ models\footnote{$N=1$ domain
walls are discussed in details in \cite{lyth4}}.

\section{Conclusion}

In this work we briefly went over axion story and two important
issues which are \begin{itemize}

\item the limits on axion models coming from the bound on the value
of isocurvature fluctuations and
\item the domain walls problem
\end{itemize}
Both issues seem to be very restrictive and provide a good test on
the viability of any axion model. One can construct reasonable
supersymmetric models which may safely avoid these problems. The
domain walls may be either diluted if PQ transition occurs prior
to the inflation, or one can break $Z_N$ symmetry by a tiny bit,
making one of the vacua more preferable then the others. One can
construct $Z_1$ models, which are domain walls problem free.
Isocurvature fluctuations problem is solved in a model with time
dependent PQ scale. More extended discussion of supersymmetric
axion models is given in \cite{ourpaper}.

\vskip 0.5in

{\bf Acknowledgments.}

I would like to thank M. Dine for many illuminating discussions,
XII Baksan Valley School for inviting me to give this talk. This
work was supported by  RFBR grant 02-02-17379.

\end{document}